# Women in Physics in the United Kingdom: A Review of Recent Policy and Initiatives

Sally Jordan[1, a)], Sarah Bakewell[2], Holly Jane Campbell[3], Josie Coltman[4], Wendy Sadler[5] and Chethana Setty[1]

[1]*School of Physical Sciences, The Open University, Milton Keynes, MK7 6AA, United Kingdom.*
[2]*Institute of Physics, 37 Caledonian Road, London, N1 9BU, United Kingdom.*
[3]*Cruachan Superconductors Ltd., 21 Stable Grove, Paisley, PA1 2DR, United Kingdom.*
[4]*AWE, Aldermaston, Reading, Berkshire, RG7 4PR, United Kingdom.*
[5]*School of Physics and Astronomy, Queen's Buildings North Building, Cardiff University, The Parade, Cardiff, CF24 3AA, United Kingdom.*

[a)] Corresponding author: sally.jordan@open.ac.uk

**Abstract.** Across the United Kingdom, initiatives designed to increase the participation and outcomes for women in physics continue, working with children of various ages as well as with adults. Improvements have been achieved by a combination of these initiatives and an accompanying strengthening of policy, but significant gender imbalances remain.

## PARTICIPATION OF GIRLS AND WOMEN IN PHYSICS

In the two years since the last report from the United Kingdom (UK) to the International Conference on Women in Physics [1], the slow improvement in the participation of women in physics has continued. However, the number of girls choosing to study physics at age 16 remains significantly lower than the number of boys choosing to study physics at the same age. There are variations between the education systems in the four nations of the United Kingdom (England, Scotland, Wales and Northern Ireland) and, while the qualifications usually studied post-16 in England, Wales and Northern Ireland are Advanced levels (known as A-levels), in Scotland students usually study a larger number of "Highers", with some going on to the more selective "Advanced Highers". The situation is further complicated by inconsistencies in the way data are collected, something the Institute of Physics (IOP) has asked the UK Government to address [2]. However, an analysis conducted in 2020 [3], shows the stark gender differences across all four nations in the numbers choosing physics at age 16. For example, in England and Wales in 2020, physics was the second most popular A-level choice for boys, but the 15th most popular choice for girls, and we have no reason to believe that this situation has improved significantly in more recent years.

**TABLE 1**. Percentage of boys and girls choosing to study physics at age 16 [3]. Note that In the UK, it is recognized that gender is no longer a binary concept. However, much historic data is only available for girls/women and boys/men.

| UK Country (date of data) | Qualification | % of boys choosing physics | % of girls choosing physics |
|---|---|---|---|
| England (2020) | A-level physics | 8.6% | 2.6% |
| Wales (2019) | A-level physics | 9% | 2% |
| Northern Ireland (2020) | A-level physics | 2.64% | 1.02% |
| Scotland (2019) | Higher physics | 72.5% | 27.5% |

The number of female undergraduate students across the UK has increased from 2765 (20% of the total population of physics undergraduates) in 2011/12 to 4780 (26% of the total population) in 2021/22 [4]. The same proportion (31%) of male and female students achieved a first class honors degree, while a slightly higher proportion of female students achieved an upper second class degree (26% compared with 24%). The number of female students on doctorate programs increased from 705 (23% of the total population) in 2010/11 to 985 (25% of the total population) in 2018/19 [5].

There has been a steady increase in the number and proportion of female academic staff in physics departments, with an increase from just 55 female full professors in 2012/13 to 135 in 2020/21 [6]. This means that in 2012/13, less than 8% of full professors in physics departments were women but by 2020/21, approaching 14% of the total were women. This is a pleasing improvement, but also highlights that there is still much room for improvement.

## POLICY AND INFLUENCING

Equality, diversity and inclusion is enshrined as a principle in the IOP strategy for both 2020-24 [7] and 2024-29 [8], with one of the four fundamental principles of the 2024-29 strategy being "Physics must welcome, include and reflect all parts of our diverse society". This principle reflects the wishes of the wider IOP membership and, in addition to being a moral imperative and a legal duty, is founded on the knowledge that diverse organizations outperform those that are not diverse [9]. The positioning of the principle in IOP strategy is important because it drives what is expected of physics educators and employers across the UK, for example by way of the new Physics Inclusion Award (discussed below) and a revised Degree Accreditation Framework which requires university physics departments to have a clear commitment to equality, diversity and inclusion if they are to offer IOP-accredited degree programs [10].

More generally, a strengthening of policy and opportunities for governmental influence have brought a potential for significant improvement at a systemic level. UK Research and Innovation (the government-sponsored group which brings together the seven disciplinary research councils and other research-related bodies) has objectives designed to create a more inclusive research and innovation culture [11]. Equality, diversity and inclusion is firmly established as a priority in most schools, universities and employers of physicists, and gender-inclusive policies, recognizing women's health issues such as menstruation and the menopause, are being introduced more widely. Unfortunately, workplace bullying and harassment is still reported frequently. Awareness training and bystander training are increasingly employed in an attempt to address this problem.

The introduction by the UK Government of shared parental leave from 2015 [12] has supported the career development and retention of female physicists. The introduction in 2017 of gender pay gap reporting [13], compulsory for all organizations with over 250 employees, has highlighted pay inequities and is encouraging a focus on reducing the pay gap. The introduction and subsequent impact of these policies illustrates how important it is for physicists to use every opportunity to influence wider policy and practice. The IOP has recently contributed to both the House of Commons Science and Technology Committee inquiry into Diversity in STEM [2] and the All-Party Parliamentary Group consultation on Equity in the STEM Workforce [14].

## ATHENA SWAN, PROJECT JUNO AND THE PHYSICS INCLUSION AWARD

Since the Athena Swan Scheme was introduced in the UK in 2005 and the launch of the Institute of Physics' Project Juno (the flagship gender equality award for university physics departments and related organizations) in 2007, the prospects for women in physics have improved markedly. A review of the Athena Swan scheme has led to the publication of transformed Athena Swan Charter principles in November 2022 [15]. Meanwhile, following extensive consultation, the IOP's highly respected Project Juno was retired in 2023 and superseded by the new, broader, Physics Inclusion Award from 2024 [16].

## INITIATIVES TO IMPROVE PARTICIPATION AND OUTCOMES

Many initiatives exist with the aim of improving participation and outcomes for girls and women in physics, at all levels, as illustrated by the following examples.

The IOP's Limit Less campaign [3] seeks to encourage more girls to study physics post-16 by influencing those who advise them (parents, carers, teachers and media), focusing on the wide range of careers that are open to those with a physics-based degree or apprenticeship.

In higher education, the annual Conference for Undergraduate Women and Non-binary Physicists (UK and Ireland) (CUWiP+) [17] brings women and non-binary undergraduates together, to encourage and support them with their personal and professional development. Before the 2020 conference, 32% of attendees said they felt confident enough to apply for postgraduate studies; after the conference, the number rose to 53%. Meanwhile, the Bell Burnell Graduate Scholarship Fund [18], established in 2018 following a generous gift from Dame Jocelyn Bell Burnell, provides financial assistance to enable students from under-represented groups in physics to complete a doctoral research degree.

In the workplace, cultural change is supported through programs such as Women in Science and Engineering (WISE) Ten Steps [19]. Women in Work (WiW) summits [20] were introduced into the UK from 2023, with the aim of prioritizing women's health and creating a future where businesses embrace gender equity, acknowledging the competitive advantage that this brings. The Daphne Jackson Trust [21] and other organizations run returner programs to support those who wish to return to their career after a break.

## DISCUSSION OF REMAINING CHALLENGES

The UK Government's Gender Equality Roadmap [22] recognizes that women take on more unpaid work than men, form the majority of informal carers, and on average do around 60% more cooking, cleaning and childcare than men. This is a significant challenge for women's participation in the workforce and their career choices and pathways.

An analysis of all the applicants to the Bell Burnell Graduate Scholarship in 2022 and 2023 found that a large proportion of them were in multiple under-represented groups [18]. Overall, 80% of applicants were women, 56% had disabilities, 56% came from financially disadvantaged backgrounds, 26% were LGBTQ+ and 46% came from ethnic minority groups. However, it is the intersections between the different categories that highlights the need to be mindful of the intersectional challenges that many girls and women in physics in the UK face, as we seek to improve the participation and outcomes for all girls and women in physics.

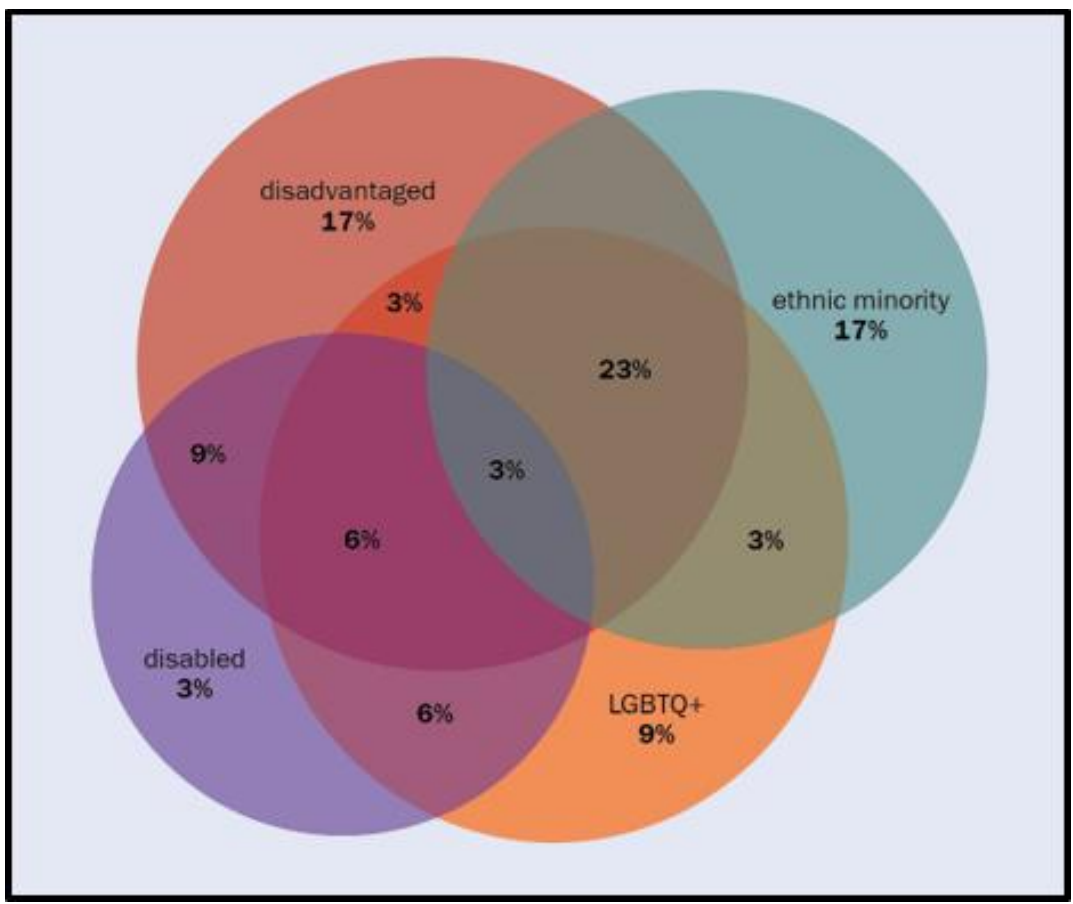

**FIGURE 1.** Analysis of all 2022 and 2023 applicants to the Bell Burnell Graduate Scholarship Fund. Figure Courtesy of IOP Publishing, from data supplied by Helen Gleeson [18]